# Magnetic anomalies in nanocrystalline $Ca_3CoRhO_6$, a geometrically frustrated spin-chain compound

*Niharika Mohapatra, Kartik K Iyer, B.A. Chalke, and E.V. Sampathkumaran*[*]
*Tata Institute of Fundamental Research, Homi Bhabha Road, Colaba, Mumbai – 400005, India.*

**Abstract**
We have investigated the magnetic behavior of the nano crystals, synthesized by high-energy ball-milling, for a well-known geometrically frustrated spin-chain system, $Ca_3CoRhO_6$, and compared its magnetic characteristics with those of the bulk form by measuring *ac* and *dc* magnetization. The features attributable to the onset of 'partially disordered antiferromagnetism' (characterizing the bulk form) are not seen in the magnetization data of the nano particles; the magnetic moment at high fields in the very low temperature range in the magnetically ordered state gets relatively enhanced in the nano particles. It appears that the ferromagnetic intrachain interaction, judged by the sign of paramagnetic Curie temperature, is preserved in the nano particles. These trends are opposite to those seen in the nano particles of $Ca_3Co_2O_6$. However, the complex spin-dynamics as evidenced by large frequency dependence of ac susceptibility is retained in the nano particles as well. Thus, there are some similarities and dissimilarities between the properties of the nano particles and those of the bulk. We believe that these findings would be useful to understand correlation lengths deciding various properties of geometrical frustration and/or spin-chain phenomena.

1. **Introduction**

The compounds of the type $(Ca,Sr)_3XYO_6$ (where X and/or Y = a magnetic-moment containing ion) crystallizing in $K_4CdCl_6$-derived rhombohedral structure have provided an unique opportunity to probe geometrically frustrated magnetism and spin-chain behavior in a single family [1]. In this family, apart from $Ca_3Co_2O_6$, the compound $Ca_3CoRhO_6$ [2] attracted considerable attention [3-17]. This compound exhibits one magnetic transition near ($T_1=$) 90 K and the other near ($T_2=$) 30 K. On the basis of neutron diffraction studies [4], it was inferred that, in the temperature ($T$) range between $T_1$ and $T_2$, partially disordered antiferromagnetism (PDA) occurs in which two out of three ferromagnetic chains are antiferromagnetically coupled while the third one does not order magnetically; at $T_2$, the incoherent chain freezes with a yet-to-be understood complex spin-dynamics as evidenced [6] by a huge frequency dependence of $ac$ susceptibility ($\chi$). It has been proposed [4] that this compound can be an example for non-equilibrium one-dimensional Ising model. There was a controversy [5, 9] surrounding valence and spin states of Co (occupying distorted trigonal prismatic coordination site) and Rh (occupying distorted octahedral site). It is now established [11, 13, and 14] that Co is in the divalent, high-spin $d^7$ spin configuration, and Rh is in the tetravalent, low-spin $d^5$ configuration. Thus, the metallic ions at both these sites are magnetic. While this situation is different from that of $Ca_3Co_2O_6$ (in which Co at both the sites are trivalent with one site being non-magnetic), the properties of both the compounds are similar in many ways [8]. Interestingly, a giant orbital moment has been found [13, 14] in the compound under investigation. This compound is also characterized by large thermoelectric power, offering a new route for the search for thermoelectric materials [7]. The magnetic properties of this compound have been found to be insensitive to applications of positive and negative pressure [16], though a small disturbance of Ca site appears to disfavor PDA formation. Recently, in addition to the transitions at $T_1$ and $T_2$, there was a microscopic experimental evidence [17] for the existence of another characteristic temperature well above $T_1$, attributable to incipient magnetic order. Thus, this compound is an interesting geometrically frustrated spin-chain compound.

Recently, we investigated the nano form of $Ca_3Co_2O_6$ synthesized by high-energy ball-milling. We found [18] that, barring very low temperature ($<< T_2$ of 10 K) isothermal magnetization ($M$) behavior, all the features including the ones attributable to PDA magnetic structure are retained in the nano form. Therefore, we considered it worthwhile to probe the magnetic properties of the nano form of the interesting compound under investigation. With this motivation, we have synthesized the nano particles of this compound by high-energy ball-milling under identical conditions as those for $Ca_3Co_2O_6$ and studied its magnetic behavior by $dc$ and $ac$ magnetization.

2. **Experimental details**

The polycrystalline samples in the bulk form (called '***A***') have been prepared as in Ref. 6. The x-ray diffraction (*XRD*) (Cu $K_\alpha$) patterns confirmed single phase nature of the compound (within the detection limit of 2%). The ball-milling conditions (employing Fritsch pulverisette-7 premium line) have been stated in Ref. 18. A small portion of the specimen after milling for 2½ h (called '***B***') was collected for magnetic studies to see how the properties evolve with milling time. The characterization by scanning electron microscope (SEM), energy dispersive x-ray analysis (EDXA), and transmission electron microscope (TEM, Tecnai 200 kV), were performed for the specimen milled for 8 hours (called '*C*'). The experimental details with respect to *dc* and *ac* magnetization (*M*) have been described in Ref. 18.

## 3. Results and discussions

In figure 1, we show the *XRD* patterns for all the specimens. We are able to index all the diffraction lines in terms of the expected structure and there is no evidence for any structural change or for any additional phase in the milled specimens within the sensitivity of the diffractometer. This is further supported by a satisfactory Reitveld fitting as shown in figure 1. We do not find any change in the lattice constants within the limits of experimental error. A reduction in the particle size due to milling is evident from the broadening of the diffraction patterns, as demonstrated in the inset of figure 1. An upper limit for the average particle size could be inferred from the width of the most intense line after subtracting instrumental line-broadening employing Debye-Scherrer formula and it turns out to be about *70* and *25* nm for B and C respectively. The SEM images, shown in figure 2a, reveal a texture attributable to significant agglomeration of the nanoparticles; the rod-shape of the isolated particles is also visible from this image. In order get a better idea of nano particles, we have carried out ultrasonification in alcohol and obtained the bright-field TEM images. These images thus obtained (see figure 2b) confirm the existence of rods with dimensions in the nanometer range. High-resolution TEM images (figure 2c and 2d) provide evidence for the crystalline nature of the nano specimens, as the signature of well-defined planes could be observed. The selected area electron diffraction pattern (see figure 2e) confirm proper structure of the compound in the nano form; the bright spots along the diffraction rings reveal that the rods are highly textured. Finally, energy dispersive x-ray analysis confirmed the stoichiometry of the nano particles.

In figure 3, the T-dependence of *dc* magnetization obtained in a field of 5 kOe is plotted for all the three specimens for field-cooled (FC from 300 K) and zero-field-cooled (ZFC) conditions of measurements. The following behavior of the bulk form has been reported at several places cited in this paper (see, for instance, Ref. 6): High temperature Curie-Weiss behavior (225 – 300 K), a broad peak around 125 K due to spin-chains, and features at $T_1$ and $T_2$ due to magnetic transitions, and the bifurcation of ZFC and FC curves. Let us now compare these features with those for the milled specimens (figure 3). We note that the features in *M/H* are to some extent retained for specimen *B* as though milling time is not adequate to attain nanoparticles. However, for *C*, the transition near 90 K is apparently washed out and we could detect the 30K-transition only. To understand the behavior above 90 K, we had to take the data with 50 kOe to suppress possible interference from the traces of a ferromagnetic impurity introduced while handling (*see also below*). We show the data in the insets of figure 3. A deviation from the Curie-Weiss region (see the insets of figure 3) persists in the milled samples, though the broad peak around 125 K in $\chi(T)$ plot vanishes for *C*. The effective moment ($\mu_{eff}$) and the paramagnetic Curie temperature ($\theta_p$) obtained from the Curie-Weiss region are: 5.4 $\mu_B$/formula-unit and ~140 K for *A*, 5.0 $\mu_B$/formula-unit and ~130 K for *B*, and 4.8 $\mu_B$/formula-unit and ~65 K for *C*, respectively. These trends in these parameters indicate that, while there could be a small change in the spin/valence states of Co and/or Rh with milling as indicated by $\mu_{eff}$, positive sign of $\theta_p$ implies dominance of the intrachain ferromagnetic interaction in nano particles though its strength undergoes relatively a small decrease. In contrast to this observation, in the case of $Ca_3Co_2O_6$, the sign of $\theta_p$ was found to be reversed [18]. It is worthwhile to verify this systematics by extending $\chi$ studies to high temperatures, considering that the Curie-Weiss region in this study is quite narrow.

We now compare isothermal remnant magnetization ($M_{IRM}$) at two temperatures, one (62 K) in the range between $T_1$ and $T_2$ and the other (1.8 K) below $T_2$ for the three specimens (Figure 4). For this purpose, we cooled the specimen in zero-field to the desired temperature, switched on a field (say, 5 kOe) for 5 mins, switched it off and then measured magnetization as a function of time (*t*). The point to note is that the decay of $M_{IRM}$ [Ref. 6] seen in *A* is absent in the milled specimens at 62 K. Instead $M_{IRM}$ increases with *t* qualitatively differing from that of *A*. Though the origin of this increase is not clear at present, the observation indicates some degree of metastability. In any case, the data reveal that the magnetic state in the intermediate temperature range (between $T_1$ and $T_2$) has undergone a profound change after milling. With respect to the behavior at 1.8 K, however, the decay (almost logarithmic) is retained after milling, attributable to spin-glass freezing.

In order to throw more light on the origin of the change in $M_{IRM}$ behavior in the temperature range between $T_1$ and $T_2$, we have measured the isothermal *M* (Figure 5) at the two temperature ranges. The plateau at one third of saturation magnetization ($M_s$) characterizing PDA state (e.g., 62 K) [5] observed for *A* is diminished for *B* with a weak increase of the slope of the curve near 50 kOe. However, this plateau is completely absent for *C* and *M* increases gradually with *H* without any evidence for saturation. This is an evidence for the suppression of PDA state as well as high-field ferromagnetism in *C*. Consistent with the absence of high-field ferromagnetic state, the magnetic moment at high fields, say, at 120 kOe, gets reduced with milling. Interestingly, in contrast to this behavior at 62 K, as the temperature is lowered below $T_2$, say, to 5 K, there is a gradual increase in the magnitude of the magnetic moment at high fields as one goes from *A* to *C* – a trend different from that seen in $Ca_3Co_2O_6$ [18]. For instance, this value for $H$ = 120 kOe is enhanced four times for *C* compared to that for *A*. Though hysteretic nature of the *M(H)* curve is preserved indicative of a ferromagnetic component, the increase of *M* with *H* is steeper in *C*, however, without any evidence for saturation as though there is an antiferromagnetic component as well. Finally, we would like to mention that, in the paramagnetic state, even at 300 K, for initial applications of field, there is a sudden increase of *M* followed by linear behavior for the milled specimens (not shown here). The value of the saturation moment extrapolated from the high-field linear region turns out to be much less than 0.01 $\mu_B$/formula-unit. We do not attach much significance to this feature at present, as it is not uncommon that it could be due to contamination while handling nano particles [19].

In figure 6, we present the *ac* $\chi$ behavior (measured in an *ac* field of 1 Oe at four frequencies, $\nu$) for the nano specimen *C*. As in the bulk material (Ref. 6), there is a peak in the real part at about 50 K for $\nu$= 1.3 Hz, which shifts to about 70 K for $\nu$= 1.333 kHz somewhat similar to that observed for the specimen *A*. Corresponding features are seen in the imaginary part as well, as described in Ref. 6. Thus, the huge frequency dependence is retained in the nano form as well, as though the slow spin-dynamics in this temperature range does not involve large length scales. However, there are a few major changes with respect to the behavior in *A*: (i) The signal strength is significantly enhanced compared to that of bulk form; (ii) the curves obtained at various $\nu$ do not overlap even near 200 K as though magnetic correlations persist even at such high temperatures; this can be related to Mössbauer anomalies [17] in the same temperature range proposed to arise from intrachain interactions; (iii) there is no upturn near 90 K, consistent with the suppression of the well-defined transition at $T_1$. These features are different from those observed for $Ca_3Co_2O_6$ [18]. An application of a magnetic field of 30 kOe completely suppresses the features, unlike in *A* [Ref.6], which means that the magnetically ordered state for the nano particle is different from that of *A* in a subtle manner.

### 4. Conclusions

We have probed the magnetic behavior of the nano crystallites of $Ca_3CoRhO_6$ synthesized by high-energy ball-milling and compared and contrasted with those of the bulk form. Major findings are: Characteristic features due to 'partially-disordered antiferromagnetism' in the range 30 to 90 K and spin-chain magnetism (a broad peak in $\chi(T)$ around 125 K), observed for bulk form, are smeared out in the *dc* magnetization data of the nano particles, though the low-temperature transition (near 30 K) and associated features seem to be preserved. However, magnetic correlations seem to extend to much higher temperatures as indicated by the frequency-dependence of *ac* $\chi$ and the value of the magnetic moment at high fields in the magnetically ordered state ($<<$ 30 K) is significantly enhanced with respect to that for the bulk form. In many respects, the observed changes for this compound, when the particle size is reduced to nano form, are different from those found for the nano form of $Ca_3Co_2O_6$. It is thus interesting that the two 'exotic' compounds in the same family with broadly similar characteristics in the bulk form behave differently in the nano form. While the precise of this difference is not clear, it is of interest to investigate whether the magnetic correlation lengths for $Ca_3CoRhO_6$ are much smaller compared to that for $Ca_3Co_2O_6$. Possibly, the existence of a moment-carrying ion (Rh) at the octahedral site in the former may also be responsible for this difference. A careful comparative neutron diffraction studies for both the compounds in the bulk and nano forms would be desirable in this regard.

**Acknowledgements**

One of us (EVS) would like to thank N.R. Selvi Jawaharlal Nehru Center for Advanced Scientific Research, Bangalore, India, for SEM data.

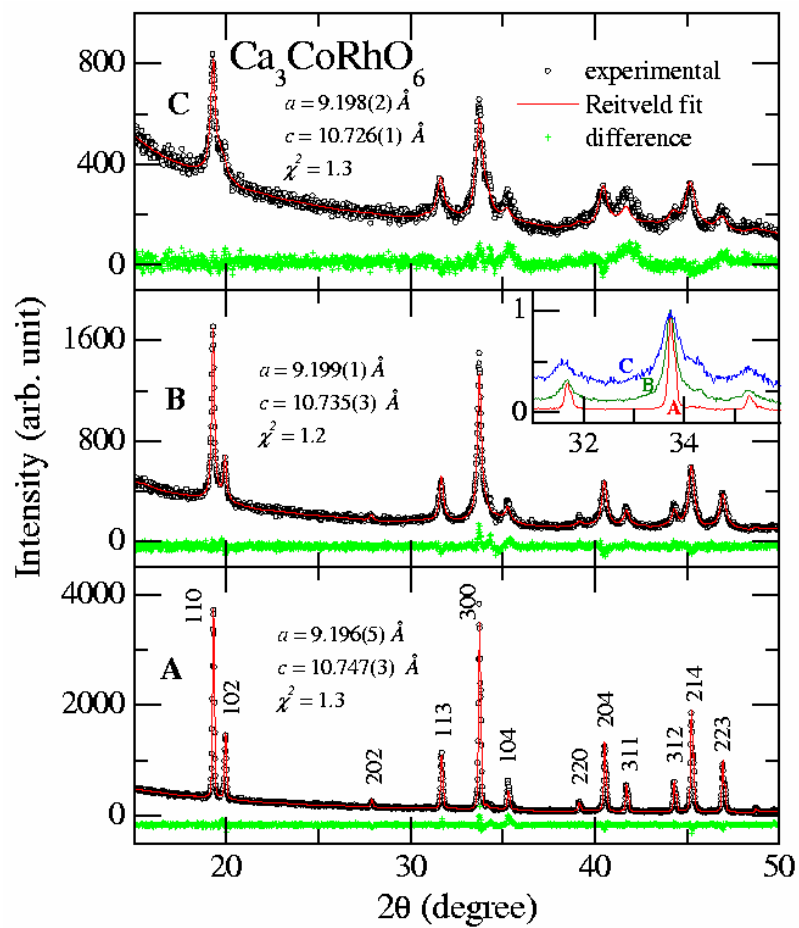

Figure 1:
(color online) X-ray diffraction pattern (Cu K$_\alpha$) of bulk (called '*A*') and the ball-milled specimens (*B*: 2½ h; *C*: 8 h) of $Ca_3CoRhO_6$. The lattice constants, the Reitveld fit, and the difference between experimental and fit-spectra are shown. In the inset, the patterns around the most intense peak are plotted after normalizing to respective peak heights to show line-broadening with milling.

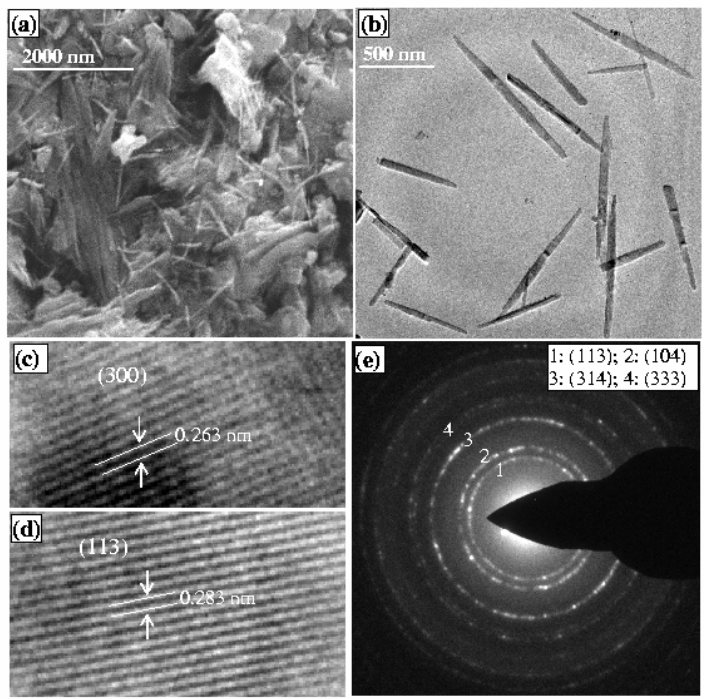

Figure 2:
(color online) **(a)** SEM images, **(b)** TEM images of nano rods, **(c)** and **(d)** high-resolution TEM showing lattice planes, and **(e)** selected area diffraction pattern obtained by TEM with indexing of innermost diffraction rings, for the ball-milled specimen, *C*, of $Ca_3CoRhO_6$.

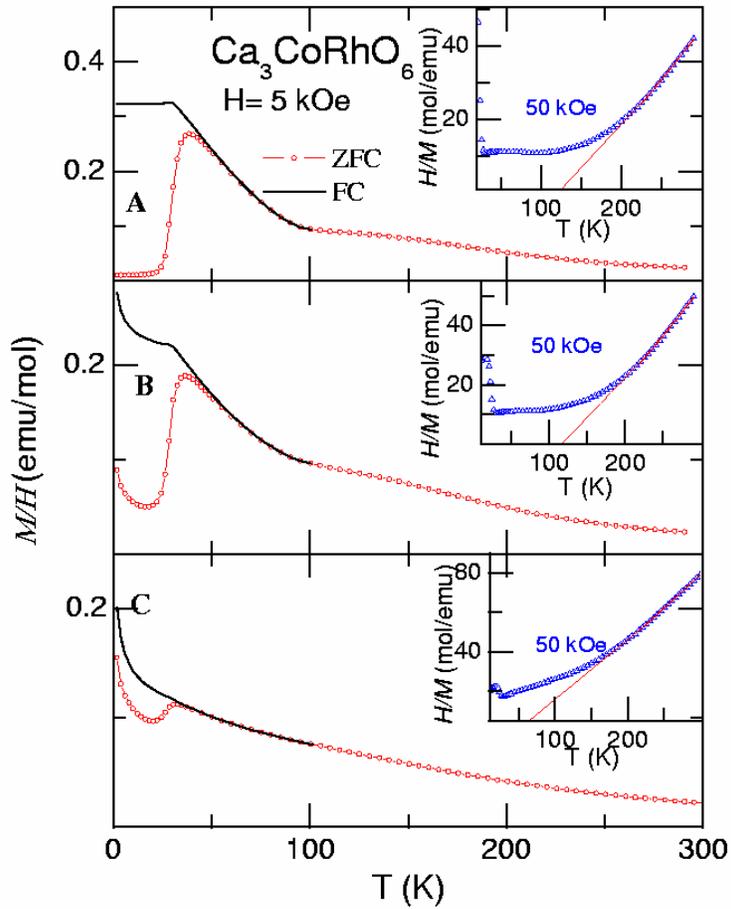

Figure 3:
(color online) Magnetization (**M**) divided by magnetic field (*H*), measured in the presence of 5 kOe for the milled samples of $Ca_3CoRhO_6$. The data for the bulk specimen are also shown for comparison. In the insets, inverse susceptibility obtained in a field of 50 kOe is plotted and the straight lines in these cases are obtained by Curie-Weiss fitting of the data above 225 K.

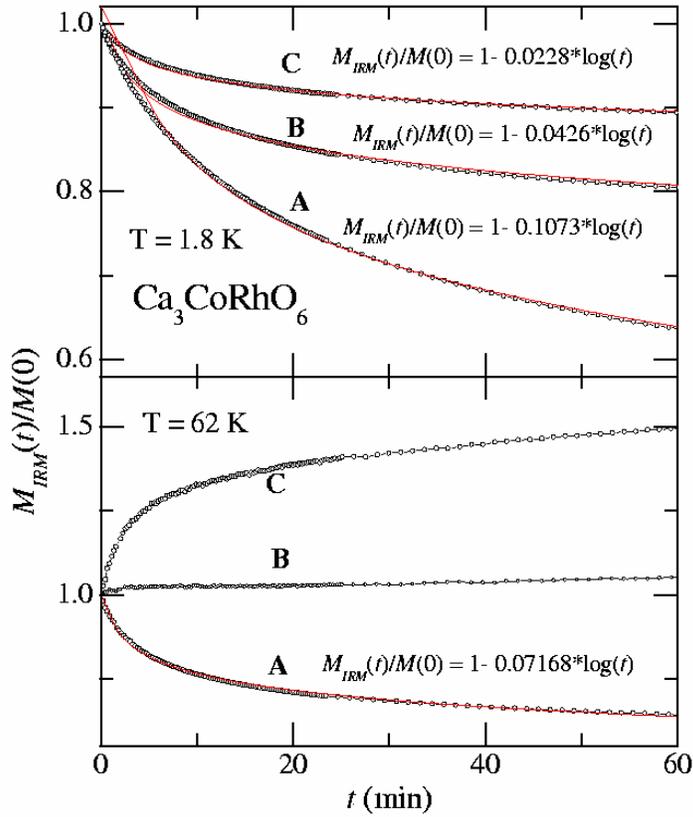

Figure 4:
(color online) Isothermal remnant magnetization behavior of bulk and milled specimens of $Ca_3CoRhO_6$, obtained as described in the text. The continuous lines represent the fit to the functions mentioned in the figures.

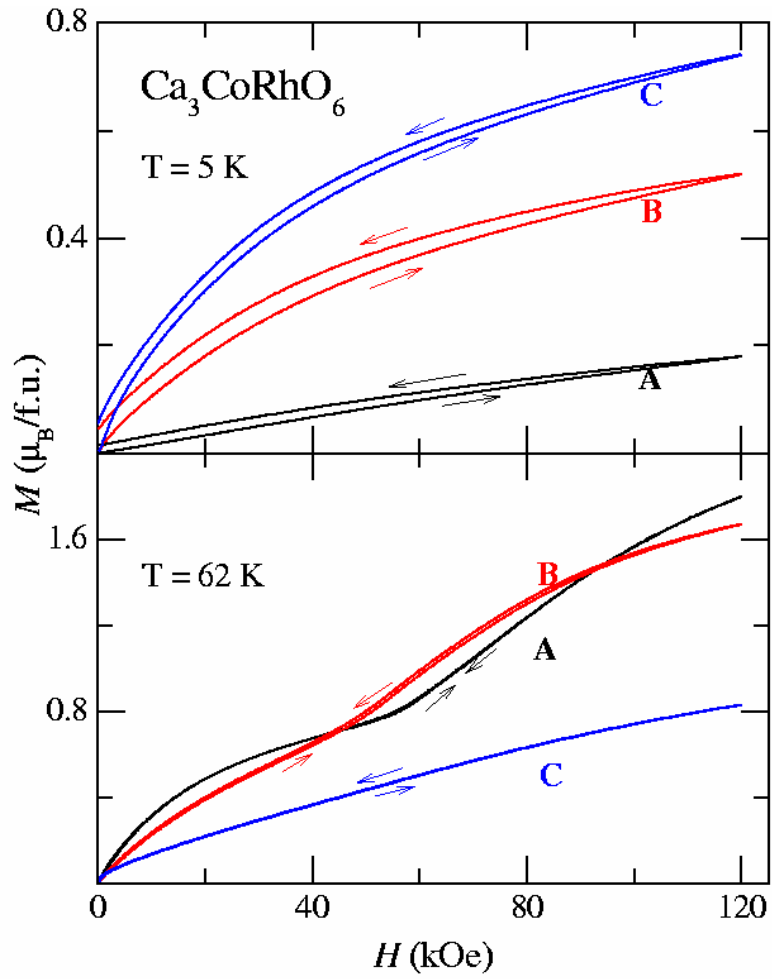

Figure 5:
(color online) Isothermal magnetization at 1.8 and 62 K for the bulk and ball-milled specimens of $Ca_3CoRhO_6$.

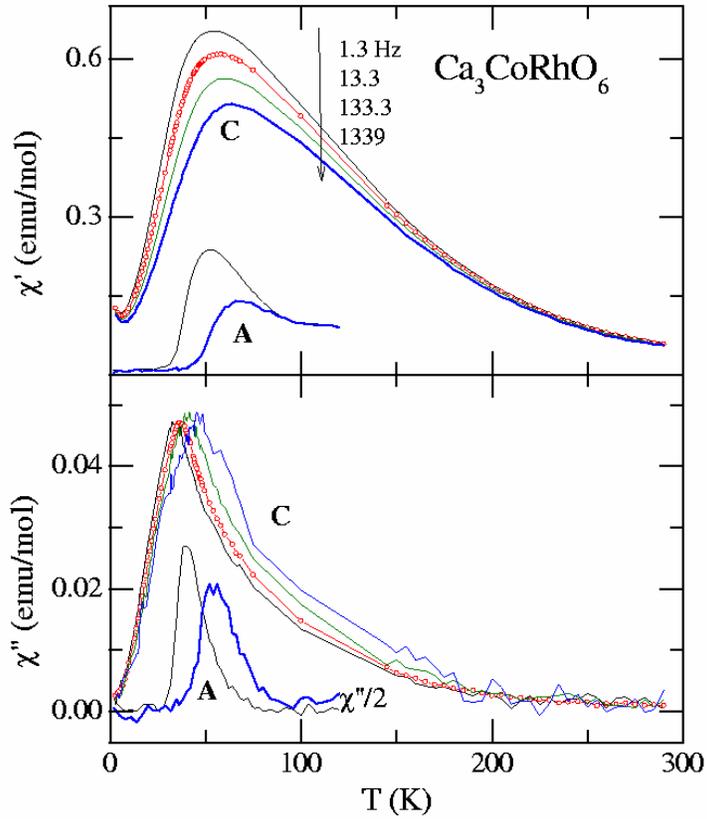

Figure 6:
(color online) Real ($\chi'$) and imaginary ($\chi''$) parts of *ac* susceptibility measured at various frequencies (with a *ac* field of 1 Oe) for the nano particles (*C*) of $Ca_3CoRhO_6$. A line through the data points only are shown, except for one curve in which case the data points are included. The curves at two extreme frequencies are shown for the bulk specimen as well for comparison.